\documentclass[preprint,tightenlines,showpacs,nofootinbib,aps,prc]{revtex4} 

\usepackage{graphicx}
\usepackage{bm}
\usepackage{url,multirow,xcolor}

\begin{document}

\title{ Two decay paths for calculation of nuclear matrix element of neutrinoless double-beta decay using quasiparticle random-phase approximation }

\author{J.\ Terasaki}
\affiliation{Division of Physics and Center for Computational Sciences, University of Tsukuba, Tsukuba 305-8577, Japan}

\begin{abstract} 
It is possible to employ virtual decay paths, including two-particle transfer, to calculate the nuclear matrix element of neutrinoless double-beta decay under the closure approximation, in addition to the true double-beta path. In the quasiparticle random-phase approximation (QRPA) approach, it is necessary to introduce the product wave functions of the like-particle and proton-neutron QRPA ground states, for  achieving consistency between the calculations of the true and virtual paths. Using these different paths, the problem of whether or not these two methods give equivalent nuclear matrix elements (NME) is investigated. It is found that the two results are inequivalent, resulting from the different many-body correlations included in the two QRPA methods, i.e., the use of the product wave functions alone is not sufficient. The author proposes introduction of the proton-neutron pairing interaction with an adequate strength in the double-beta-path method, which carries less many-body correlations without this supplemental interaction, for obtaining the NME equivalent to that of the two-particle-transfer-path method. The validity of the proposed modified approach is examined. 
\end{abstract}

%
\maketitle
\section{\label{sec:introduction}Introduction}
The nuclear matrix element (NME) of neutrinoless double-beta ($0\nu\beta\beta$) decay has been studied intensively by many theoretical research groups, who have been inspired by the attempts initiated by several experimental groups to observe $0\nu\beta\beta$ decay \cite{Ume08,Arn08,EXO14,Ack13,Gan13} (experiments already started). 
Exploitation of the $0\nu\beta\beta$ decay is one of the few available methods for determining the neutrino mass scale. In this approach, it is necessary to obtain reliable values for the NME and phase-space factor in order to determine the effective neutrino mass accurately, together with the experimental decay probability. However, it is difficult to establish the NME reliably, because the predicted $0\nu\beta\beta$ NMEs cannot be confirmed by any other methods. This difficulty arises because the transition operator is a neutrino-exchange interaction in medium, which has never been used for the investigation of other nuclear physics problems. In addition, an uncertainty is introduced through approximation of the nuclear wave function, as the lightest candidate parent nucleus used in experiment is $^{48}$Ca \cite{Ume08}. 
A well-known problem exists in that the values of the $0\nu\beta\beta$ NME are distributed within a factor range of 2--3 depending on the method (see, e.g., \cite{Fae12J}). This uncertainty may affect the design of next-generation detectors significantly, as can be conjectured from discussions on experimental sensitivity (see, e.g., \cite{Ver12}). 
Therefore, improvement of the NME calculation is an urgent challenge for theoretical nuclear physics. The author emphasizes that study of the many-body problem is essential in order to solve the problem of discrepancy between methods.

The approach to the NME calculation that utilizes the quasiparticle random-phase approximation (QRPA), which is employed in this paper, has been improved over a number of decades, in terms of deformation of the nuclear density \cite{Rad93}, evaluation of the finite nucleon-size effect (see, e.g., \cite{Sim08}),  the tensor term in the transition operator (e.g., \cite{Sim08}), the renormalized QRPA (\cite{Rod02,Pac03} and references therein), several effective-operator methods (e.g., \cite{Kor07}), and other factors. Remarkable progress has been made recently, in that a few groups have independently performed QRPA calculations (\cite{Fan15} and references therein), with the converged results with respect to the dimension of the wave-function space, and relatively similar $0\nu\beta\beta$ NMEs were obtained within a range significantly smaller than the abovementioned factor of 2--3.
 
Recently, the author has made additional progress by modifying the QRPA approach for calculation of the NME of $0\nu\beta\beta$ decay in terms of the overlap calculation of the intermediate states of the decay \cite{Ter13}. In the QRPA approach, the intermediate states obtained on the basis of the initial and final states are not identical; therefore, this overlap calculation is not trivial. 
A major modification involves the explicit use of the QRPA ground-state wave function, which is defined as vacuum for QRPA bosons. Hence, it has been found that the product of the normalization factors of the QRPA ground states is almost 2 for the  $^{150}$Nd$\rightarrow^{150}$Sm decay, which has a significant effect in terms of NME reduction \cite{Ter15}. The author's NME is close to that of the conventional QRPA approach \cite{Fan11, Fan15}; however, the modified method has an important physical implication. In the conventional approach, the proton-neutron (pn) pairing interaction is introduced so as to reproduce the experimental NME or decay probability of the two-neutrino double-beta ($2\nu\beta\beta$) decay, and the pnQRPA solutions are close to the breaking point of the pnQRPA as a result of the strong pn pairing interaction. The advantage of the modified QRPA approach is that such a strong pn pairing interaction is unnecessary; therefore, the QRPA approach is, in fact, a better approximation than has previously been recognized.

Another new aspect of the modified QRPA approach is the use of a virtual decay path, e.g.,~a two-neutron removal followed by a two-proton addition, in order to calculate the NME of $0\nu\beta\beta$ decay under the closure approximation \cite{Vog10,Ter13,Bro14}. The like-particle (lp)QRPA is the main method for obtaining intermediate states on  this virtual path. On the other hand, the pnQRPA is the primary method for obtaining those on the conventional double-beta  path. It is, however, necessary to extend the QRPA ground states to product wave functions of the lp and pn QRPA ground states in order to unify the ground-state wave functions used in the two methods. This extension implies that the product of the normalization factors of the lpQRPA ground states is also included in the overlap of the pnQRPA states. Therefore, the NME obtained via the double-beta-path calculation is also reduced, and the strong pn pairing interaction can be avoided. 

This paper is the third major step in the author's examination and improvement of the manner in which the QRPA is applied to the calculation of the $\beta\beta$ NME. 
The problem investigated in this paper is whether or not the previously reported modified QRPA approach \cite{Ter15} yields equivalent results for the two decay paths, namely, the double-beta and two-particle-transfer paths. Note that the author previously reported the result of a small-space calculation indicating that these two paths yield close NMEs \cite{Ter15}. However, this equivalence is a property of exact nuclear wave functions, and it is not trivial for approximate wave functions. In this paper, we investigate whether this equivalence is obtained exactly in the QRPA. The fundamental purpose of the study of this problem is deepening our understanding of the QRPA approach, and such research is necessary because the uncertainties of the methods must be removed individually in order to solve the method-based NME discrepancy problem. 

Section \ref{sec:preparation} presents basic equations for calculating the NME. In Sec.~\ref{sec:cross_term}, the problem of QRPA equivalence between the two decay paths is investigated both analytically and numerically. The results of the calculation of the $0\nu\beta\beta$ NME using the double-beta path and the $2\nu\beta\beta$ NME are shown in Sec.~\ref{sec:calculation}. The consistency between the two different path calculations is also discussed in this section, along with a method for satisfying the equivalence of those calculations and a justification of the proposed method. Section \ref{sec:conclusion} contains the conclusion and a brief discussion of future prospects. 

\section{\label{sec:preparation}Preparation}
The probability of the $0\nu\beta\beta$ decay is given by
\begin{eqnarray}
1/T_{0\nu}(gs\rightarrow gs) = |M^{(0\nu)}|^2 G_{01} \left( \langle m_\nu \rangle /m_e \right)^2, 
\label{eq:0vbb_decay_prob}
\end{eqnarray}
(see, e.g., \cite{Doi85}) where $T_{0\nu}(gs\rightarrow gs)$ is the half-life from the ground to ground states, $G_{01}$ is the phase-space factor, $\langle m_\nu \rangle$ is the effective neutrino mass, and $m_e$ is the electron mass. We use the nuclear matrix element consisting of only the Gamow-Teller (double $\sigma\tau$) and Fermi (double $\tau$) terms for simplicity
\begin{eqnarray}
M^{(0\nu)} &\simeq& 
\sum_{K\pi}\sum_{a_F^{K\pi} a_I^{K\pi}}\sum_{\alpha\alpha^\prime:p}\sum_{\beta\beta^\prime:n}
\langle -\alpha\alpha^\prime | h_+(r_{12},\bar{E}) \Big\{ -\bm{\sigma}(1)\cdot\bm{\sigma}(2) 
+g_V^2/g_A^2 \Big\} \tau^+(1) \tau^+(2) | \beta -\beta^\prime \rangle \nonumber\\
&&\langle F| c_\beta c_{-\alpha}^\dagger |a_F^{K\pi}\rangle 
\langle a_F^{K\pi} | a_I^{K\pi}\rangle 
\langle a_I^{K\pi}| c_{\alpha^\prime}^\dagger c_{-\beta^\prime} |I\rangle, 
\label{eq:0vbb_NME}
\end{eqnarray}
where $\alpha$ and $\alpha^\prime$ ($\beta$ and $\beta^\prime$) denote proton (neutron) single-particle states, and their combinations are constrained with respect to the $z$-component of the angular momentum by 
\begin{eqnarray}
j_{\beta^\prime}^z +j_{\alpha^\prime}^z = K, \ 
j_\beta^z +j_\alpha^z = K, 
\label{eq:constraints_jz_alpha_beta}
\end{eqnarray}
and with respect to parity by 
\begin{eqnarray}
\pi_{\beta^\prime} \pi_{\alpha^\prime} = \pi, \ 
\pi_\beta \pi_\alpha = \pi, 
\label{eq:constraints_pi_alpha_beta}
\end{eqnarray}
with the quantum number $K$ and parity $\pi$ of the intermediate state. Axial and parity symmetries are assumed for the nuclei considered in this paper. When $-j_\alpha^z$ is included in the single-particle label, we note $-\alpha$. The creation and annihilation operators of those single-particle states are denoted by $c$ and $c^\dagger$, respectively, and the $a_I^{K\pi}$ and $a_F^{K\pi}$ labels denote the intermediate states obtained on the basis of the initial ($I$) and final ($F$) states, respectively. The $h_+(r_{12},\bar{E})$ function is the neutrino potential \cite{Ter15}, with $r_{12}$ and $\bar{E}$ being the inter-nucleon distance and average intermediate-state energy, respectively. $\bm{\sigma}$ and $\tau^+$ are the spin-Pauli and charge-change (neutron to proton) operators, respectively. Arguments 1 and 2 distinguish the two particles on which the operators act, and the $g_V$ and $g_A$ constants are the vector current and axial-vector current couplings, respectively. 

In the proposed QRPA approach, the $I$ and $F$ states are product states of the lp and pn QRPA ground states such that
\begin{eqnarray}
|I\rangle &=& \frac{1}{{\cal N}_{pn,I} {\cal N}_{lp,I}} \prod_{K\pi} 
\exp{ \big[ v_{pn,I}^{K\pi} \big] } \exp{ \big[ v_{lp,I}^{K\pi} \big] } |i\rangle, \nonumber \\
|F\rangle &=& \frac{1}{{\cal N}_{pn,F} {\cal N}_{lp,F}} \prod_{K\pi} 
\exp{ \big[ v_{pn,F}^{K\pi} \big] } \exp{ \big[ v_{lp,F}^{K\pi} \big] } |f\rangle, 
\label{eq:I_F}
\end{eqnarray}
\begin{eqnarray}
v_{pn,I}^{K\pi} &=& \sum_{\mu\nu\mu^\prime\nu^\prime}
C_{\mu\nu,-\mu^\prime-\nu^\prime}^{pnI,K\pi}
a_\mu^{i\dagger} a_\nu^{i\dagger} a_{-\mu^\prime}^{i\dagger} a_{-\nu^\prime}^{i\dagger}, 
\ \mu \textrm{ and } -\mu^\prime: \textrm{proton};
\ \nu \textrm{ and } -\nu^\prime: \textrm{neutron}, \nonumber\\
v_{lp,I}^{K\pi} &=& \sum_{\mu\nu\mu^\prime\nu^\prime}
C_{\mu\nu,-\mu^\prime-\nu^\prime}^{lpI,K\pi}
a_\mu^{i\dagger} a_\nu^{i\dagger} a_{-\mu^\prime}^{i\dagger} a_{-\nu^\prime}^{i\dagger}, 
\ \mu \textrm{ and } \nu \ (-\mu^\prime \textrm{ and } -\nu^\prime): \textrm{lp's}, \nonumber\\
v_{pn,F}^{K\pi} &=& \sum_{\mu\nu\mu^\prime\nu^\prime}
C_{\mu\nu,-\mu^\prime-\nu^\prime}^{pnF,K\pi}
a_\mu^{f\dagger} a_\nu^{f\dagger} a_{-\mu^\prime}^{f\dagger} a_{-\nu^\prime}^{f\dagger}, 
\ \mu \textrm{ and } -\mu^\prime: \textrm{proton};
\ \nu \textrm{ and } -\nu^\prime: \textrm{neutron}, \nonumber\\
v_{lp,F}^{K\pi} &=& \sum_{\mu\nu\mu^\prime\nu^\prime}
C_{\mu\nu,-\mu^\prime-\nu^\prime}^{lpF,K\pi}
a_\mu^{f\dagger} a_\nu^{f\dagger} a_{-\mu^\prime}^{f\dagger} a_{-\nu^\prime}^{f\dagger}, 
\ \mu \textrm{ and } \nu \ (-\mu^\prime \textrm{ and } -\nu^\prime): \textrm{lp's}, 
\label{eq:v}
\end{eqnarray}
where $\{a_\mu^{i\dagger}, a_\mu^i\}$ denote the creation and annihilation operators of the canonical quasiparticle state $\mu$ associated with the initial Hartree-Fock-Bogoliubov (HFB) ground state $|i\rangle$, respectively, and $\{a_\mu^{f\dagger}, a_\mu^f\}$ denote the creation and annihilation operators associated with the final HFB ground state $|f\rangle$, respectively. 
Further, ${\cal N}_{pn,I}$, ${\cal N}_{lp,I}$, ${\cal N}_{pn,F}$, and ${\cal N}_{lp,F}$ are the normalization factors of the QRPA ground states defined by
\begin{eqnarray}
{\cal N}_{pn,I}^2 &=& \prod_{K\pi} \langle i | \exp[v^{K\pi\dagger}_{pn,I}] 
\exp[v^{K\pi}_{pn,I}] | i \rangle , \nonumber \\
{\cal N}_{lp,I}^2 &=& \prod_{K\pi} \langle i | \exp[v^{K\pi\dagger}_{lp,I}] 
\exp[v^{K\pi}_{lp,I}] | i \rangle , \nonumber \\
{\cal N}_{pn,F}^2 &=& \prod_{K\pi} \langle f | \exp[v^{K\pi\dagger}_{pn,F}] 
\exp[v^{K\pi}_{pn,F}] | f \rangle , \nonumber \\
{\cal N}_{lp,F}^2 &=& \prod_{K\pi} \langle f | \exp[v^{K\pi\dagger}_{lp,F}] 
\exp[v^{K\pi}_{lp,F}] | f \rangle . \label{eq:normalization_factor}
\end{eqnarray} 
The coefficients in Eq.~(\ref{eq:v}) are referred to as the correlation coefficients (see, e.g., \cite{Ull72}). Using the quasiboson approximation which neglects exchange terms, these coefficients are obtained as
\begin{eqnarray}
C_{\mu\nu,-\mu^\prime-\nu^\prime}^{pnI,K\pi} &=& 
\frac{1}{1+\delta_{K0}} \sum_a Y^{pnI,K\pi,a\ast}_{-\mu^\prime-\nu^\prime}
\left( \frac{1}{ X^{pnI,K\pi\ast} } \right)_{a,\mu\nu} , \nonumber \\
C_{\mu\nu,-\mu^\prime-\nu^\prime}^{lpI,K\pi} &=& 
\frac{1}{1+\delta_{K0}} \sum_a Y^{lpI,K\pi,a\ast}_{-\mu^\prime-\nu^\prime}
\left( \frac{1}{ X^{lpI,K\pi\ast} } \right)_{a,\mu\nu} , \nonumber \\
C_{\mu\nu,-\mu^\prime-\nu^\prime}^{pnF,K\pi} &=& 
\frac{1}{1+\delta_{K0}} \sum_a Y^{pnF,K\pi,a\ast}_{-\mu^\prime-\nu^\prime}
\left( \frac{1}{ X^{pnF,K\pi\ast} } \right)_{a,\mu\nu} , \nonumber \\
C_{\mu\nu,-\mu^\prime-\nu^\prime}^{lpF,K\pi} &=& 
\frac{1}{1+\delta_{K0}} \sum_a Y^{lpF,K\pi,a\ast}_{-\mu^\prime-\nu^\prime}
\left( \frac{1}{ X^{lpF,K\pi\ast} } \right)_{a,\mu\nu} . \label{eq:C}
\end{eqnarray}
The $X$ and $Y$ symbols are used to represent the forward and backward amplitudes of the QRPA solutions, respectively. As seen from these equations, we use four kinds of QRPA solutions, which are distinguished by $\{pn, lp\}$ and $\{I, F\}$. A QRPA solution is then specified by, e.g., $\{pnI,K\pi,a\}$. $X^{pn,I,K\pi}$ represents the matrix for which the row (column) number corresponds to $\mu\nu$ ($a$). The creation operators of the pnQRPA states are given by
\begin{eqnarray}
O^{pnI,K\pi\dagger}_a &=& \sum_{\mu\nu} \left( X^{pnI,K\pi,a}_{\mu\nu} a^{i\dagger}_\mu
a^{i\dagger}_\nu - Y^{pnI,K\pi,a}_{-\mu-\nu} a^{i}_{-\nu} a^{i}_{-\mu} \right), \nonumber \\
O^{pnF,K\pi\dagger}_a &=& \sum_{\mu\nu} \left( X^{pnF,K\pi,a}_{\mu\nu} a^{f\dagger}_\mu
a^{f\dagger}_\nu - Y^{pnF,K\pi,a}_{-\mu-\nu} a^{f}_{-\nu} a^{f}_{-\mu} \right), 
 (\mu: \textrm{proton}, \nu: \textrm{neutron}). \label{eq:Opndagger} 
\end{eqnarray}
The intermediate states in Eq.~(\ref{eq:0vbb_NME}) can be expressed as
\begin{eqnarray}
|a^{K\pi}_I \rangle &=& O^{pnI,K\pi\dagger}_a |I\rangle , \nonumber \\
|a^{K\pi}_F \rangle &=& O^{pnF,K\pi\dagger}_a |F\rangle . \label{eq:intermediate_states_pn}
\end{eqnarray}

We also note the NME calculated using a two-particle-transfer path
\begin{eqnarray}
M^{(0\nu)}_{2p2n} &\simeq& \sum_{K\pi}\sum_{m_{F}^{K\pi} m_{I}^{K\pi}}\sum_{\alpha\alpha^\prime:p}\sum_{\beta\beta^\prime:n}
\langle -\alpha\alpha^\prime | h_+(r_{12},\bar{E}) \Big\{ -\bm{\sigma}(1)\cdot\bm{\sigma}(2) 
+g_V^2/g_A^2 \Big\} \tau^+(1) \tau^+(2) | \beta -\beta^\prime \rangle \nonumber\\
&&\langle F| c_{-\alpha}^\dagger c_{\alpha^\prime}^\dagger |m_F^{K\pi}\rangle 
\langle m_F^{K\pi} | m_I^{K\pi}\rangle 
\langle m_I^{K\pi}| c_\beta c_{-\beta^\prime} |I\rangle,
\label{eq:0vbb_NME_2p2n} 
\end{eqnarray}
\begin{eqnarray}
|m^{K\pi}_I \rangle &=& O^{lpI,K\pi\dagger}_m |I\rangle , \nonumber \\
|m^{K\pi}_F \rangle &=& O^{lpF,K\pi\dagger}_m |F\rangle , \label{eq:intermediate_states_lp}
\end{eqnarray}
\begin{eqnarray}
O^{lpI,K\pi\dagger}_m &=& \sum_{\mu<\nu} \left( X^{lpI,K\pi,m}_{\mu\nu} a^{i\dagger}_\mu
a^{i\dagger}_\nu - Y^{lpI,K\pi,m}_{-\mu-\nu} a^{i}_{-\nu} a^{i}_{-\mu} \right), \nonumber \\
O^{lpF,K\pi\dagger}_m &=& \sum_{\mu<\nu} \left( X^{lpF,K\pi,m}_{\mu\nu} a^{f\dagger}_\mu
a^{f\dagger}_\nu - Y^{lpF,K\pi,m}_{-\mu-\nu} a^{f}_{-\nu} a^{f}_{-\mu} \right), 
(\mu \textrm{ and } \nu : \textrm{lp's})\label{eq:Olpdagger} 
\end{eqnarray}
where $\mu<\nu$ indicates that $(\mu,\nu)$ is an ordered pair of lp's. The $a^{K\pi}_I$ and $a^{K\pi}_F$ labels are used for the pnQRPA states created on $|I\rangle $ and $|F\rangle$, respectively, and 
$m^{K\pi}_I$ and $m^{K\pi}_F$ are used for the corresponding lpQRPA states. 
For details on the lpQRPA and pnQRPA calculations, see Refs.~\cite{Ter05,Ter10} and Ref.~\cite{Ben02}, respectively.

\section{\label{sec:cross_term}Charge-change transition matrix element}
In this section, the matrix element of the transition density in $M^{(0\nu)}$, i.e., Eq.~(\ref{eq:0vbb_NME}), is investigated for the extension of the QRPA ground states to the product states. In the computation of $\langle a^{K\pi}_I | c^{\dagger}_{\alpha^\prime} c_{-\beta^\prime} | I \rangle$ and $\langle F| c_\beta c_{-\alpha}^\dagger |a_F^{K\pi}\rangle$, $\{ c^{i\dagger}_\alpha, c^i_\alpha \}$ and
$\{ c^{f\dagger}_\alpha, c^f_\alpha \}$ are used, respectively, for the single particles. The choice of these two single-particle bases is useful for the calculation of the transition densities, and the matrix elements of the two-body transition operator in Eq.~(\ref{eq:0vbb_NME}) are calculated using these bases. 
The two-body matrix elements are calculated using the coordinate mesh; thus, no additional calculation arises for this part.
Using Eq.~(\ref{eq:I_F}), we can state 
\begin{eqnarray}
\lefteqn{ \langle a^{K\pi}_I | c^{i\dagger}_{\alpha^\prime} c^i_{-\beta^\prime} | I \rangle }
\nonumber\\
&&=\frac{1}{ {\cal N}_{lp,I}^2 {\cal N}_{pn,I}^2 } \langle i | \prod_{K\pi} 
\exp[v^{K\pi\dagger}_{lp,I}] \exp[v^{K\pi\dagger}_{pn,I}] O^{pnI,K\pi}_a c^{i\dagger}_{\alpha^\prime}
c^i_{-\beta^\prime} 
\exp[v^{K\pi}_{pn,I}] \exp[v^{K\pi}_{lp,I}] | i \rangle . \label{eq:transition_density_me_pn_lp}
\end{eqnarray}
The operators associated with the lpQRPA and pnQRPA are commutable under the QRPA, because there is no coupling between the two QRPA Hamiltonians. Thus, Eq.~(\ref{eq:transition_density_me_pn_lp}) leads to 
\begin{eqnarray}
\langle a^{K\pi}_I | c^{i\dagger}_{\alpha^\prime} c^i_{-\beta^\prime} | I \rangle
&\simeq& \frac{1}{ {\cal N}_{pn,I}^2 } \langle i | \prod_{K\pi}  \exp[v^{K\pi\dagger}_{pn,I}] O^{pnI,K\pi}_a c^{i\dagger}_{\alpha^\prime} c^i_{-\beta^\prime} \exp[v^{K\pi}_{pn,I}] | i \rangle, 
\nonumber \\
&=& \langle a^{K\pi}_{pn,I} | c^{i\dagger}_{\alpha^\prime} c^i_{-\beta^\prime} | I_{pn} \rangle ,
\label{eq:transition_density_me_reduced}
\end{eqnarray}
where the following definitions of the conventional pnQRPA states (the lpQRPA correlations are not included) are employed:
\begin{eqnarray}
| I_{pn} \rangle = \frac{1}{ {\cal N}_{pn,I} } \prod_{K\pi} \exp[v^{K\pi}_{pn,I}] |i\rangle,
\nonumber \\
| a^{K\pi}_{pn,I}\rangle = O^{pnI,K\pi\dagger}_a | I_{pn}\rangle. 
\label{eq:not-extended_pnQRPA_states}
\end{eqnarray}
Note that $v^{K\pi}_{lp,I}$ consists of products of operators associated with the lpQRPA, and 
$c^{i\dagger}_{\alpha^\prime} c^i_{-\beta^\prime}$ can be expressed as a linear combination of 
$O^{pnI,K\pi\dagger}_a$ and $O^{pnI,K\pi}_a$. 
Apparently, analogous equations also hold for $\langle F | c^f_{\beta} c^{f\dagger}_{-\alpha} | a^{K\pi}_F \rangle$. It has already been found that the effects of $v^{K\pi}_{lp,I}$ on the unnormalized overlap matrix elements are very small \cite{Ter15}. Thus, the lpQRPA correlations are incorporated into the NME expression [Eq. (\ref{eq:0vbb_NME})] by the normalization factors of the lpQRPA ground states only, in an approximate manner. 
$M^{(0\nu)}_{2p2n}$ includes the pnQRPA correlations in the normalization factors of the pnQRPA only, again in an approximate manner, and those normalization factors are common to the two NMEs. Therefore, the two NMEs differ, because there are differences in the components of the interactions contributing to the two NMEs; see the argument presented in Sec.~VII of Ref.~\cite{Ter15}. In other words, introduction of the product QRPA ground states is not sufficient for establishing equivalence for the two different path calculations of the NME. 

We investigate whether or not Eq.~(\ref{eq:transition_density_me_reduced}) is a good approximation. To obtain a computable equation, we  
first expand the unnormalized factor of $\langle a^{K\pi}_I | c^{i\dagger}_{\alpha^\prime} c^i_{-\beta^\prime} | I \rangle$ with respect to $v^{K\pi}_{lp,I}$ and truncate it at the first order. Next, we expand this first-order term with respect to $v^{K\pi}_{pn,I}$ and leave the zeroth-order term only. 
This approximation yields
\begin{eqnarray}
\langle a^{K\pi}_I | c^{i\dagger}_{\alpha^\prime} c^i_{-\beta^\prime} | I \rangle
\simeq \frac{1}{ {\cal N}_{lp,I}^2 } \bigg(
\langle a^{K\pi}_{pn,I} | c^{i\dagger}_{\alpha^\prime} c^i_{-\beta^\prime} | I_{pn} \rangle 
+\frac{1}{ {\cal N}_{pn,I}^2 } \prod_{K\pi} 
\langle i | O^{pnI,K\pi}_a c^{i\dagger}_{\alpha^\prime} c^i_{-\beta^\prime} 
v^{K\pi}_{lp,I} | i \rangle
 \bigg). 
\label{eq:pn_transition_density_truncated} 
\end{eqnarray}
In this equation, $\langle i | v^{K\pi\dagger}_{lp,I} O^{pnI,K\pi}_a 
c^{i\dagger}_{\alpha^\prime} c^i_{-\beta^\prime} | i \rangle $ is also neglected, because it is smaller than 
$\langle i | O^{pnI,K\pi}_a c^{i\dagger}_{\alpha^\prime} c^i_{-\beta^\prime} 
v^{K\pi}_{lp,I} | i \rangle$
by a factor of $Y$. After some analytical calculation, we obtain
\begin{eqnarray} 
\langle i | O^{pnI,K\pi}_{a} c^{i\dagger}_{\alpha} c^i_{-\beta} v^{K\pi}_{lp,I} | i \rangle 
= \sum_{\kappa < \lambda} X^{pnI,K\pi,a\ast}_{\kappa\lambda} \check{C}^{lpI,K\pi}_{\kappa\lambda,-\alpha-\beta} , \label{eq:cross_term_including_check_C}
\end{eqnarray}
\begin{eqnarray}
\check{C}^{lpI,K\pi}_{\kappa\lambda,-\alpha-\beta} &=& \frac{j^{iz}_\alpha}{|j^{iz}_\alpha|} v^i_\alpha u^i_\beta
\bigg( -C^{lpI,K\pi}_{\lambda-\beta,-\alpha\kappa} 
+C^{lpI,K\pi}_{\lambda-\beta,\kappa-\alpha}
+C^{lpI,K\pi}_{-\beta\lambda,-\alpha\kappa}
-C^{lpI,K\pi}_{-\beta\lambda,\kappa-\alpha} \nonumber\\
&&+C^{lpI,K\pi}_{\kappa-\alpha,\lambda-\beta}
-C^{lpI,K\pi}_{\kappa-\alpha,-\beta\lambda}
-C^{lpI,K\pi}_{-\alpha\kappa,\lambda-\beta}
+C^{lpI,K\pi}_{-\alpha\kappa,-\beta\lambda} \bigg). 
\label{eq:check_C}
\end{eqnarray}
The $u^i_\alpha$ and $v^i_\alpha$ factors are defined by the transformation 
\begin{eqnarray}
c^{i\dagger}_\alpha = u^i_\alpha a^{i\dagger}_\alpha +\frac{j^{iz}_\alpha}{|j^{iz}_\alpha|}
v^i_\alpha a^i_{-\alpha}, 
\end{eqnarray}
between the canonical and canonical-quasiparticle bases, 
where the $j^{iz}_\alpha/|j^{iz}_\alpha|$ phase is our phase convention associated with time reversal. 
There are selection rules for the components of Eq.~(\ref{eq:cross_term_including_check_C}); for example, for the term proportional to $X^{pnI,K\pi,a\ast}_{\kappa\lambda} C^{lpI,K\pi}_{\lambda-\beta,-\alpha\kappa}$, we have
\begin{eqnarray}
&&j^{iz}_\kappa + j^{iz}_\lambda = K, \ \pi^i_\kappa \pi^i_\lambda = \pi, \nonumber \\
&&j^{iz}_\lambda +j^{iz}_{-\beta} = K, \ \pi^i_\lambda \pi^i_{-\beta} = \pi, \nonumber \\
&&j^{iz}_{-\alpha}+j^{iz}_\kappa = -K, \ \pi^i_{-\alpha} \pi^i_\kappa = \pi. \label{eq:selection_rule_cross_term}
\end{eqnarray}
These equations imply that one more set of conditions must be satisfied than in the case of the direct term; one can compare this condition with analogous condition for, e.g., the 
$X^{pnI,K\pi,a\ast}_{\kappa\lambda} C^{pnI,K\pi}_{\kappa\lambda,-\alpha-\beta}$ component included in the first term of the right-hand side of Eq.~(\ref{eq:pn_transition_density_truncated}) implicitly. Therefore the ``cross term'' of the pn and lp QRPA (\ref{eq:cross_term_including_check_C}) has significantly less components than the analogous direct terms. 
It is speculated that the ``cross term'' is significantly smaller than the first term of the right-hand side of Eq.~(\ref{eq:pn_transition_density_truncated}) contained in the QRPA as a result of its smaller number of components. This speculation is based on the assumption that the ``cross term'' does not contain a significantly greater number of large terms than the analogous direct terms.  

Similar to Eq.~(\ref{eq:cross_term_including_check_C}), we derive
\begin{eqnarray} 
\langle f | v^{K\pi}_{lp,F} c^f_\beta c^{f\dagger}_{-\alpha} O^{pnF,K\pi\dagger}_{a} | f \rangle 
= \sum_{\kappa < \lambda} X^{pnF,K\pi,a\ast}_{\kappa\lambda} \check{C}^{lpF,K\pi}_{\kappa\lambda,-\alpha-\beta}, \label{eq:cross_term_including_check_C_F}
\end{eqnarray}
\begin{eqnarray}
\check{C}^{lpF,K\pi}_{\kappa\lambda,-\alpha-\beta} &=& \frac{ j^{fz}_{-\beta} }{|j^{fz}_\beta|} v^f_\beta u^f_\alpha
\bigg( -C^{lpF,K\pi}_{\kappa-\alpha,\lambda-\beta} 
+C^{lpF,K\pi}_{\kappa-\alpha,-\beta\lambda}
+C^{lpF,K\pi}_{-\alpha\kappa,\lambda-\beta}
-C^{lpF,K\pi}_{-\alpha\kappa,-\beta\lambda} \nonumber\\
&&+C^{lpF,K\pi}_{\lambda-\beta,-\alpha\kappa}
-C^{lpF,K\pi}_{\lambda-\beta,\kappa-\alpha}
-C^{lpF,K\pi}_{-\beta\lambda,-\alpha\kappa}
+C^{lpF,K\pi}_{-\beta\lambda,\kappa-\alpha} \bigg). 
\label{eq:check_C_F}
\end{eqnarray}
The calculations of the ``cross terms'', Eqs.~(\ref{eq:cross_term_including_check_C}) and (\ref{eq:cross_term_including_check_C_F}), are computationally costly and challenging for parallel computation. For the calculation and verification, see Appendix \ref{app:cross_term_comp_check}. 

The numerical calculation of the NME using Eq.~(\ref{eq:pn_transition_density_truncated}) and the corresponding calculation including Eq.~(\ref{eq:cross_term_including_check_C_F}) were performed for $(K\pi)$ = $(2+)$ in  $^{150}$Nd $\rightarrow$ $^{150}$Sm using  the same setup as that used to calculate $M^{(0\nu)}_{2p2n}$ in Ref.~\cite{Ter15}. That is, the HFB solutions were common to both cases, and the cutoff parameters were chosen in such a way that the dimensions of the QRPA Hamiltonian matrix and the $0\nu\beta\beta$-transition-operator matrix were close to those used in the calculation of $M^{(0\nu)}_{2p2n}$.
The truncated $v^{K\pi}_{lp,I}$ and $v^{K\pi}_{lp,F}$ in Ref.~\cite{Ter15} were also used to calculate the overlaps of the pnQRPA intermediate states. These operators were created in order to reproduce the correlation energies of the ground states, and $v^{K\pi}_{pn,I}=v^{K\pi}_{pn,F}=0$ was employed in the overlap calculation because the major QRPA solutions with the large backward amplitudes were those of the lpQRPA. For more details, see Ref.~\cite{Ter15}. Note that $v^{K\pi}_{pn,I}=v^{K\pi}_{pn,F}=0$ is not assumed in the above general discussion.   
A $M^{(0\nu)}$ component of $(K\pi)=(2+)$ was obtained to be 0.3712 in this test calculation. 
This value is identical to that obtained without the ``cross terms'' [Eqs. (\ref{eq:cross_term_including_check_C}) and (\ref{eq:cross_term_including_check_C_F})] within the shown order. A calculation using all the lpQRPA solutions with the same $(K\pi)$ value was also performed in order to construct another pair of $v_{lp,I}^{K\pi}$ and $v_{lp,F}^{K\pi}$, and the corresponding NME value was 0.3710. Thus, assuming commutability of the operators associated with the different QRPAs is a very good approximation in this test. 

This result indicates the necessity of the intermediate-state projector consisting of multi-phonon states [see Eq.~(7) in Ref.~\cite{Ter13}]. However, those multi-phonon states do not contribute to the transition density in the QRPA if the exchange terms are neglected. It is also speculated that the contribution from those multi-phonon states is also negligible if the exchange terms are included, based on the numerical calculation conducted in this section. (Note that the numerical confirmation is too computationally costly to perform.) Thus, dynamical extension of the QRPA is the appropriate means of obtaining equivalence for the two different path calculations. Nevertheless, an alternative method for obtaining equivalence in the QRPA is considered in the next section. 

\section{\label{sec:calculation}Calculation of nuclear matrix element using double-beta path}
\subsection{\label{subsec:0vbb_NME} $\bm{0\nu\beta\beta}$ nuclear matrix element}
The $M^{(0\nu)}$ calculation was performed for $^{150}$Nd $\rightarrow$ $^{150}$Sm using  the same input as the ``cross term'' calculation in the previous section. Thus, the cutoff parameters for the two-quasiparticle spaces used in the pnQRPA calculations were chosen such that the associated dimensions were close to those of the lpQRPA calculations in Ref.~\cite{Ter15}. However, the excitation modes of $K=0$ and 1, $\pi=+$ and $-$ had spurious states in the lpQRPA, and much larger spaces were used than those for other $K\pi$'s. The cutoff parameters for those $K\pi$'s in the present calculation were chosen to yield dimensions close to that of $(K\pi)=(2+)$ in Ref.~\cite{Ter15}. 
A value of 5.324 was obtained for the $M^{(0\nu)}$. As a value of 3.604 was obtained for $M^{(0\nu)}_{2p2n}$ in Ref.~\cite{Ter15}, the result obtained in this study is 48\% larger than the earlier value. 

\begin{table}
\caption{\label{tab:three_NMEs} $0\nu\beta\beta$ NME values obtained in the present calculation for $^{150}$Nd$\rightarrow$$^{150}$Sm. $\hat{V}^\textrm{\scriptsize{pnpair}}$ was introduced with the strength equal to average lp-pairing interaction strength for the protons and neutrons (see text).}
\begin{ruledtabular}
\begin{tabular}{lc}
Method & $0\nu\beta\beta$ NME  \\
\colrule
Two-particle-transfer path, Eq.~(\ref{eq:0vbb_NME_2p2n}) & 3.604  \\
Double-beta path, Eq.~(\ref{eq:0vbb_NME}), without $\hat{V}^\textrm{\scriptsize{pnpair}}$ & 5.324 \\
Double-beta path, Eq.~(\ref{eq:0vbb_NME}), with $\hat{V}^\textrm{\scriptsize{pnpair}}$ & 3.697 \\
\end{tabular}
\end{ruledtabular}
\end{table}
\begin{figure}[t]
\includegraphics[width=7.5cm]{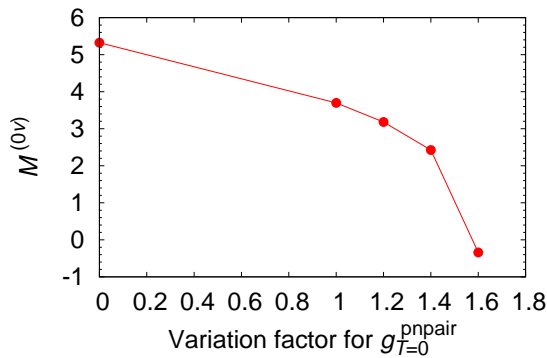}
\caption{ \label{fig:nme-dbpath_factor} (Color online) $M^{(0\nu)}$ against variation factor for   $g^\textrm{\scriptsize{pnpair}}_{T=0}$ value obtained by averaging lp-pairing interaction strengths for protons and neutrons.}
\end{figure}
A number of groups (e.g., \cite{Mus13}) have independently reported that the larger the strength of the pn pairing interaction, the smaller the obtained $M^{(0\nu)}$. Using this property, the pn pairing interaction for retrieving the equivalence of the two different path calculations can be introduced. The pnQRPA calculations are modified by this interaction, and the lpQRPA calculations are not affected. In this paper, a contact volume interaction is used, which consists of the component acting on the pn pair with $T=0$ and that with  $T=1$. This interaction can be expressed as   
\begin{eqnarray}
\hat{V}^\textrm{\scriptsize{pnpair}} &=& 
g_{T=0}^\textrm{\scriptsize{pnpair}} \delta(\bm{r}_1-\bm{r}_2) P_{S=1} P_{T=0} 
+g_{T=1}^\textrm{\scriptsize{pnpair}} \delta(\bm{r}_1-\bm{r}_2) P_{S=0} P_{T=1}, 
\label{eq:pn_pairing_interaction}
\end{eqnarray}
where $P_S$ and $P_T$ are projection operators to the spin $S$ and isospin $T$ pair states, respectively, and $g^\textrm{\scriptsize{pnpair}}_T$ denotes the interaction strength. The equation of the matrix element of this interaction used in the pnQRPA equation is shown in Appendix \ref{app:me_pn_pairing_int}. 
The $T=1$ pn pairing interaction (the interaction is not isovector) is often used to cause the Fermi term of the $2\nu\beta\beta$ NME to vanish (this term vanishes automatically if the nucleus has isospin symmetry). 
However, the $T=1$ pn pairing interaction is omitted because, without this, the Fermi term of the $2\nu\beta\beta$ NME is 7--10 \% (varying as a result of the uncertainty of $g_A$) of the entire NME in the present calculation.
Using $g^\textrm{\scriptsize{pnpair}}_{T=0}=-197.44$ and $-200.09$ MeVfm$^3$ for $^{150}$Nd and $^{150}$Sm, respectively, the $M^{(0\nu)}$ of 3.697 was obtained. These $g^\textrm{\scriptsize{pnpair}}_{T=0}$ values are averages of the lp-pairing interactions strengths for the protons and neutrons. These average values yield reasonable equivalence of the two different path calculations, although the analytical reason for this result has not been determined. 
The three NME values obtained are summarized in Tab.~\ref{tab:three_NMEs}, and 
the dependence of $M^{(0\nu)}$ on $g^\textrm{\scriptsize{pnpair}}_{T=0}$ is illustrated in Fig.~\ref{fig:nme-dbpath_factor}. The $g^\textrm{\scriptsize{pnpair}}_{T=0}$ adopted here, which corresponds to a factor of 1, is still far from the breaking point of the pnQRPA indicated by the sharply decreasing behavior observed in the figure.  

As the ground states under consideration here do not have pn pairing gaps, the pn pairing interaction is a residual interaction, i.e.,~only correlations beyond the HFB approximation are created. Thus, the many-body correlations due to this interaction reduce $M^{(0\nu)}$. Figure 7 in Ref.~\cite{Ter15} shows that the pnQRPA solutions obtained without the pn pairing interaction have less many-body correlations in terms of the backward amplitudes than the lpQRPA solutions. The pn pairing interaction in the present calculation is an effective interaction for supplementing the many-body correlations of the pnQRPA solutions. Further, this interaction is useful for the purposes of the present study, because it does not affect the lpQRPA solutions. Thus, the pn pairing interaction is employed without discussion of its physical origin. As discussed in Ref.~\cite{Ter15}, it is difficult to find the physical origin of the pn pairing interaction for nuclei far from $N=Z$.

\subsection{\label{subsec:2vbb_NME} $\bm{2\nu\beta\beta}$ nuclear matrix element}
The calculation of the $2\nu\beta\beta$ NME is an important check for theory, particularly because experimental data are available. This NME is calculated according to \cite{Doi85}
\begin{eqnarray}
M^{(2\nu)} &=& \frac{ M^{(2\nu)}_{GT} }{ \mu_0 }
-\frac{g_V^2}{g_A^2} \frac{ M^{(2\nu)}_F }{ \mu_{0F} }, \label{eq:2vbb_NME}
\end{eqnarray}
\begin{eqnarray}
\frac{M^{(2\nu)}_{GT}}{\mu_0} &=& \sum_{a_I^K,a_F^K} \frac{1}{\mu_a} \langle F | \sum_n \tau^+(n) (-)^{K} [\sigma(n)]_{-K} | a_F^K \rangle \langle a_F^K | a_I^K \rangle \langle a_I^K | \sum_{n^\prime} 
\tau^+(n^\prime) [\sigma(n^\prime)]_K | I \rangle \nonumber \\
&&\times
\Bigg\{
\begin{array}{l}
2, K=1, \\
1, K=0,
\end{array} \label{eq:2vbb_NME_GT} \\
\frac{M^{(2\nu)}_F}{\mu_{0F}} &=& \sum_{a_I,a_F} \frac{1}{\mu_a} \langle F| \sum_n \tau^+(n) | a_F \rangle
\langle a_F | a_I \rangle \langle a_I | \sum_{n^\prime} \tau^+(n^\prime) |I \rangle,
\label{eq:2vbb_NME_F} \\
\mu_a &=& \frac{1}{m_e c^2} \Big\{ E_{aK,I} -\frac{1}{2}(M_F +M_I) \Big\}.
\end{eqnarray}
The running indexes $n$ and $n^\prime$ indicate the nucleons, and $E_{aK,I}$ is the energy of the intermediate state. $M_I$ and $M_F$ are the masses of the initial and final states, respectively. The intermediate states are limited to those with positive parity and $K=0$ and 1 only for the GT term [Eq.~(\ref{eq:2vbb_NME_GT})] and $K=0$ for the Fermi term [Eq.~(\ref{eq:2vbb_NME_F})]. 

\begin{table}
\caption{\label{tab:2vbb_NME} $2\nu\beta\beta$ NME values obtained in the present calculation for $^{150}$Nd$\rightarrow$$^{150}$Sm and its components. Semi-experimental values are also shown, which were obtained from the experimental half-life \cite{Bar13}, theoretical phase-space factor \cite{Kot12}, and $g_A$. $|K|$ indicates those of the intermediate states. 
For $K\neq 0$, the NME components shown here are summations of those for $K=\pm |K|$.}
\begin{ruledtabular}
\begin{tabular}{ccccc}
$|K|$ & $M^{(2\nu)}_{GT}(|K|)/\mu_0$ & $M^{(2\nu)}_F(|K|)/\mu_{0F}$ & \multicolumn{2}{c}{$M^{(2\nu)}(|K|)$}  \\
      &                         &                         & $g_A=1.254$           & $g_A=1.000$  \\
\colrule
0     &  0.0271 & $-$0.0092 & 0.0329 & 0.0363 \\
1     &  0.0486 & 0         & 0.0486 & 0.0486 \\
\hline
      & $M^{(2\nu)}_{GT}/\mu_0$ & $M^{(2\nu)}_F/\mu_{0F}$ & \multicolumn{2}{c}{$M^{(2\nu)}$}  \\
\hline
Total &  0.0757 & $-$0.0092 & 0.0816 & 0.0849 \\
Semi-exp. & & & 0.0368 & 0.0579
\end{tabular}
\end{ruledtabular}
\end{table}
%
\begin{figure}[t]
\includegraphics[width=7.5cm]{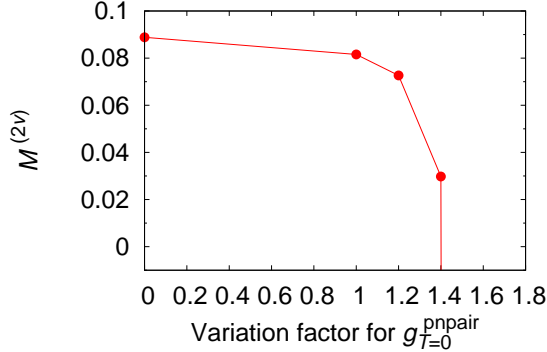}
\caption{ \label{fig:nme2v_factor} (Color online) Identical to Fig.~\ref{fig:nme-dbpath_factor}, but for $M^{(2\nu)}$.}
\end{figure}
The result of the $2\nu\beta\beta$ calculation and semi-experimental values are shown in Tab.~\ref{tab:2vbb_NME}. As mentioned above, the  Fermi-term contribution is small, and the difference due to the two $g_A$ is not large. The semi-experimental values of $M^{(2\nu)}$ obtained using the compiled data of Ref.~\cite{Bar13}, the phase-space factor of Ref.~\cite{Kot12}, and the values of $g_A$ are 0.0579 ($g_A=1.000$) and 0.0368 ($g_A=1.254$). 
Apparently, the value calculated with $g_A=1.000$ is closer to the semi-experimental value in terms of the ratio than that obtained for $g_A=1.254$.
However, the $M^{(2\nu)}$ obtained here with $g_A=1.000$ is 47 \% larger than the semi-experimental value. 
In Ref.~\cite{Ter13}, the author obtained a value of 0.06, which is very close to the semi-experimental result, via a simple hybrid estimation combining the $M^{(2\nu)}$ of Ref.~\cite{Mus13} and the product of the normalization factors of the lpQRPA ground states. This very good agreement is a coincidence. However, the discrepancy of 47 \% is still encouraging, considering that this result was obtained without phenomenology for reproducing the semi-experimental value. 

In Fig.~\ref{fig:nme2v_factor}, the $g^\textrm{\scriptsize{pnpair}}_{T=0}$ dependence of $M^{(2\nu)}$ is illustrated. As previously noted  in the literature (e.g., \cite{Mus13}), the decreasing behavior in the large-$g^\textrm{\scriptsize{pnpair}}_{T=0}$ region is sharper than that of $M^{(0\nu)}$. Again, this figure shows that the present $g^\textrm{\scriptsize{pnpair}}_{T=0}$ is in a region in which the pnQRPA provides a good approximation, indicating that the proposed method is useful. 

\subsection{\label{subsec:technical_comparison}Technical comparison with previously reported calculation}
The calculation reported in Ref.~\cite{Mus13} was performed independently of this study, using the Skyrme interaction (energy-density functional) SkM$^\ast$ \cite{Bar82} and the contact volume pairing interaction; those interactions are identical to those employed here, except for the lp-pairing strength.\footnote{The lp-pairing strengths are not available in Ref.~\cite{Mus13}; however, the principle used to determine the strength is identical in both Ref.~\cite{Mus13} and the present study. That is, the three-point formula is used to reproduce the pairing gaps deduced from the experimental masses.} Thus, a comparison to determine the technical consistency is worthwhile. For this purpose, the simplest example of $M^{(2\nu)}$ was employed, without the pn pairing interaction with $g_A=1.0$, and the new method presented here was not employed. The $M^{(2\nu)}$ value of Ref.~\cite{Mus13} was read to be 0.11, from Fig.~2 of that paper [assuming that the plot shows the values of their Eq.~(6) multiplied by $g_A^2m_ec^2$]. The initial value for deriving the corresponding value was obtained using the new method presented in this study and neglecting the Fermi term, and was found to be 0.0830. This result was multiplied by ${\cal N}_{lp,F} {\cal N}_{lp,I}=1.860$ (erasing the effect of the new approach) with an adjusting factor due to the difference in the nuclear mean radius of 1.2/1.1. The $M^{(2\nu)}$ result was then 0.168; 
this is 50 \% larger than the value reported in Ref.~\cite{Mus13}. 
After this calculation, three more changes were made to the calculation setup: The $T=1$ pn pairing interaction was introduced so as to cause the Fermi term of $M^{(2\nu)}$ to vanish; the dimensions of the two-quasiparticle spaces were adjusted to be closer to those of Ref.~\cite{Mus13}; and the same number of mesh points as in Ref.~\cite{Mus13} were used to represent the nucleon wave functions. Hence, a $M^{(2\nu)}$ value of 0.143 was obtained. 

Then, the two HFB solutions were compared, with slight differences being found. The Q-value reported in Ref.~\cite{Mus13} was 2.35 MeV, whereas the present result was 1.406 MeV (the experimental value is 3.371 MeV). The $\beta$ deformations of the HFB solutions were 0.27 (Ref.~\cite{Mus13}) and 0.279 (this study) for $^{150}$Nd, and 0.22 (Ref.~\cite{Mus13}) and 0.209 (this study) for $^{150}$Sm. The strengths of the lp pairing interaction of $^{150}$Sm were increased by 2 \%, and a Q-value of 2.268 MeV and a $\beta$ of 0.197 ($^{150}$Sm) were obtained. Using this input, a $M^{(2\nu)}$ result of 0.126 was determined; the discrepancy between the $M^{(2\nu)}$ value obtained in this study and that of Ref.~\cite{Mus13} was therefore reduced to 13 \%, although the present result remained larger than the Ref.~\cite{Mus13} value. 
An $M^{(2\nu)}$ value appreciably closer to that reported in Ref.~\cite{Mus13} was obtained by approaching the input used in that study; thus, it is concluded that there is no technical inconsistency between the techniques employed by the two groups as far as $M^{(2\nu)}$ is concerned.  
Other information obtained from this check is that an adjustment of approximately 10 \% in the NME can be obtained via a slight change in the HFB calculation. The accumulation of the differences in the other technical details can also yield a comparable difference in the calculated NME.

\section{\label{sec:conclusion}Conclusion and Prospects}
NME calculations have been performed for $^{150}$Nd $\rightarrow$ $^{150}$Sm using the double-beta path, and the consistency of the result compared with calculations using the virtual two-particle-transfer path in the QRPA has been investigated. It has been clarified that the QRPA does not yield exact equivalence for the two different path calculations. This problem has been investigated in depth, both analytically and numerically. The fundamental reason for the discrepancy is that the lpQRPA has a greater amount of many-body correlations than the pnQRPA, and the product QRPA ground state does not remove the influence of this difference because of the nature of the included transition matrix elements. The pn pairing interaction was introduced to obtain equivalence, exploiting the fact that the pnQRPA solutions are sensitive to this interaction but the lpQRPA solutions are unaffected. 
The essential concept is to carry the same degree of many-body correlations in the two calculations.
Interestingly, this method constitutes a theoretical prescription to determine the strengths of the $T=0$ pn pairing interaction. 
The QRPA solutions obtained using this modified method are still far from regions close to the breaking point of the QRPA, and the $M^{(2\nu)}$ does not differ largely from the semi-experimental data. Thus, the overall conclusion of the present study, including the previous papers of the author \cite{Ter13,Ter15}, is that the validity of the proposed method is confirmed. 
A technical check of the calculation was also performed by comparing the results with the findings of another group, which were obtained using the Skyrme interaction; no inconsistency was found. This test helped establish the reliability of the calculations. 

The possibility for basic improvement of the manner in which the QRPA is applied to the $\beta\beta$ NME seems to have been exhausted by this paper. Other possible improvements to the QRPA approach may involve use of the appropriate effective $g_A$, suitable interaction choice, appropriate tensor terms in the transition operator (as have already been introduced by other groups), consideration of the contributions of exotic matter, e.g., heavy neutrinos \cite{Doi85}, and other factors. 

The next task appears to be calculation of the NMEs of other decay instances so as to clarify the nucleus-dependent features. An important question is whether or not the calculation reliability is uniform for those instances. Another important step is dynamical extension of the (renormalized) QRPA, as mentioned before. However, the present finding, i.e., that the QRPA solutions are in the applicable region of the QRPA, implies the possibility that the effects beyond the QRPA are not as dramatic in the NME as a factor of 2$-$3. Considering the status of the method-based NME discrepancy problem, examination of this possibility is very important. 

\appendix
\section{\label{app:cross_term_comp_check}Computation and check of lpQRPA term in charge-change transition matrix element}

Equation (\ref{eq:cross_term_including_check_C}) was calculated via parallel computation using  ScaLAPACK \cite{Sca12}. That is, the multiplication of matrices $X^{pnI,K\pi\ast}$ and $\check{C}^{lpI,K\pi}$ (the matrix elements are $X^{pnI,K\pi,a\ast}_{\kappa\lambda}$ and $\check{C}^{lpI,K\pi}_{\kappa\lambda,-\alpha-\beta}$) was performed using the block-cyclic distribution. 
The calculation of $C^{lpI,K\pi}$ [Eq.~(\ref{eq:C})] was also performed using ScaLAPACK, by solving a linear equation. The subset of the matrix elements of $\{C^{lpI,K\pi}_{\kappa\lambda,-\alpha-\beta}\}$ carried by the ScaLAPACK process when the linear equation was solved differed from that necessary for the matrix multiplication of $X^{pnI,K\pi}$ and $\check{C}^{lpI,K\pi}$, as can be seen from the indexes of $C^{lpI,K\pi}$ in Eq.~(\ref{eq:check_C}). 
Note that, when a similar calculation was required in Ref.~\cite{Ter15}, for high-order terms in the expansion of the unnormalized overlap of two intermediate QRPA states with respect to the QRPA correlations, the components of those high-order terms were truncated using the occupation probabilities of the involved canonical single-particle states. A rather dramatic truncation was possible, because the effects of the QRPA correlations on the unnormalized overlap was small. Further, that calculation was performed without the ScaLAPACK, but with an ordinary parallel scheme of the message passing interface \cite{mpi15}. As Eq.~(\ref{eq:cross_term_including_check_C}) was then calculated on the basis of the initial state only, a dramatic truncation was prudently avoided. 
Thus, the subset of  $\{C^{lpI,K\pi}_{\kappa\lambda,-\alpha-\beta}\}$ was transferred from a ScaLAPACK process to all the processes, and each receiver process selected the matrix elements necessary for that process to perform the next matrix multiplication. This transference was made from all the processes; thus, this step was computationally costly. However, this additional communication is necessary in order to use the exchange terms (i.e., terms beyond the Boson approximation) without the truncation. The calculations based on the final state were performed in the same manner.  

\begin{figure}[t]
\vspace{11pt}
\includegraphics[width=6cm]{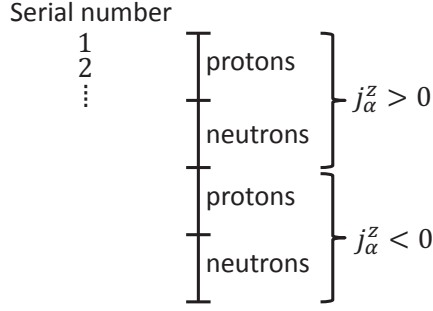}
\caption{ \label{fig:order_sp} Order of single-particle states.}
\end{figure}
\begin{figure}[t]
\vspace{11pt}
\includegraphics[width=8cm]{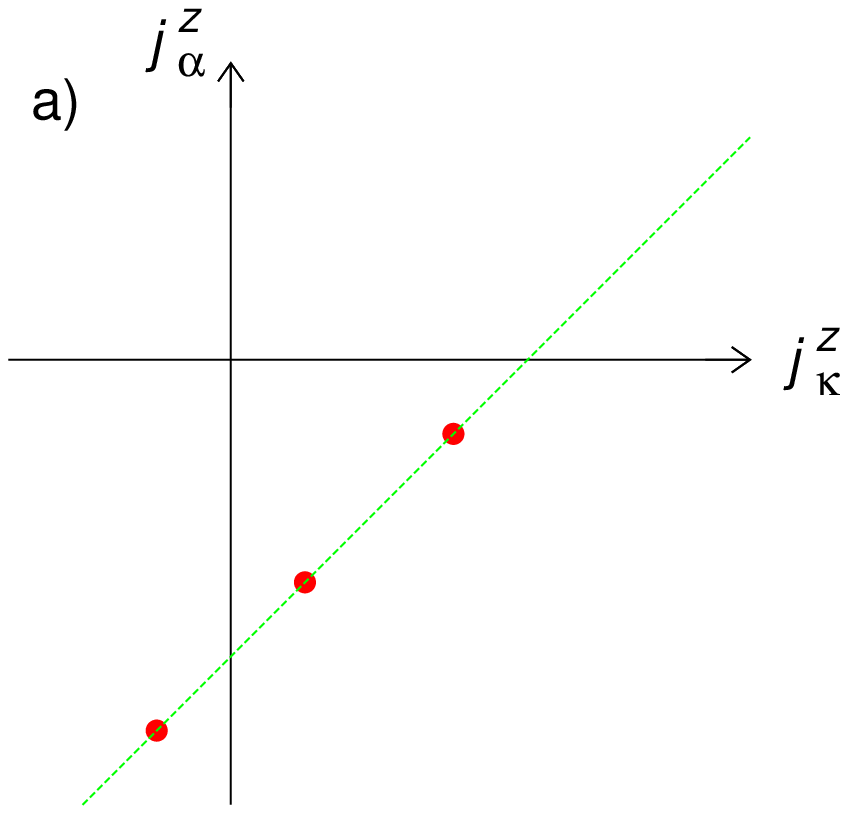}
\includegraphics[width=8cm]{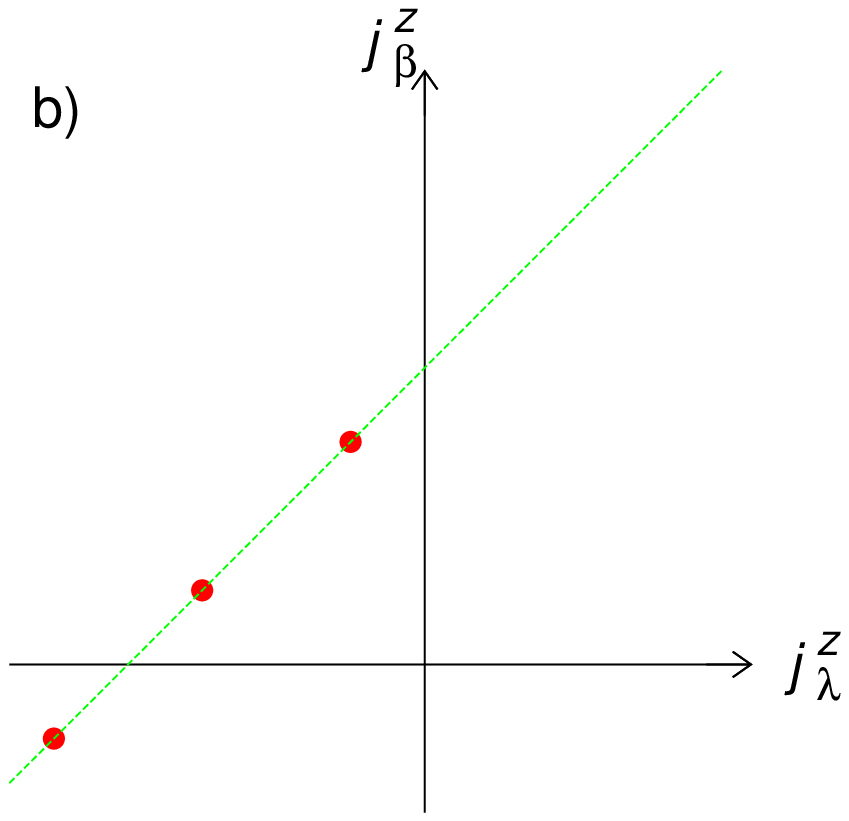}
\caption{ \label{fig:jz_alpha_kappa} (Color online) a) Possible $(j^z_\kappa,j^z_\alpha)$ for $K=2$, shown by filled circles. The dashed line is a guiding line expressing $j^z_\alpha=j^z_\kappa -K$. b) Possible $(j^z_\lambda,j^z_\beta)$ for $K=2$. The guiding line is $j^z_\beta=j^z_\lambda +K$. }
\end{figure}
A selection rule exists for the components of $\check{C}^{lpI,K\pi}_{\kappa\lambda,-\alpha-\beta}$, depending on the employed convention (other than that discussed in Sec.~\ref{sec:cross_term}), and this property can be used as a check of the computation. The canonical single-particle indexes of, e.g.,  $C^{lpF,K\pi}_{-\alpha\kappa,\lambda-\beta}$, fall under the following conditions. The $-\alpha$ and $\kappa$ states are proton states, whereas the $\lambda$ and $-\beta$ states are neutron states. From the convention employed in the definition of $X^{lpF,K\pi,a}_{-\alpha\kappa}$ and $Y^{lpF,K\pi,a}_{\lambda-\beta}$, the following orders hold:

\begin{eqnarray}
-\alpha < \kappa,\  -\lambda < \beta. \label{eq:order_condition_two_sp}
\end{eqnarray}
In addition, the conditions 
\begin{eqnarray}
j^z_{-\alpha} +j^z_\kappa &=& K, \nonumber \\
j^z_{-\lambda} +j^z_\beta &=& K, \nonumber \\
j^z_\alpha +j^z_\beta &=& K, \nonumber \\
j^z_\kappa +j^z_\lambda &=& K, \label{eq:jz_conditions_3rd_term}
\end{eqnarray}
are in effect. Therefore, 
\begin{eqnarray}
j^z_{-\alpha}>0,\  j^z_{-\lambda}>0, \label{eq:jz_larger_than_0}
\end{eqnarray}
are necessary for $K\geq 0$ (numerical calculations with $K<0$ can be omitted using the time-reversal symmetry), as a result of the order conditions (\ref{eq:order_condition_two_sp}) and the conventional order of single-particle states shown in Fig.~\ref{fig:order_sp}. 
From condition (\ref{eq:jz_larger_than_0}) and the first and second relations of Eq.~(\ref{eq:jz_conditions_3rd_term}), the possible $(j^z_\kappa,j^z_\alpha)$ and $(j^z_\lambda,j^z_\beta)$ are restricted, as illustrated in Fig.~\ref{fig:jz_alpha_kappa}. From these figures, it is apparent that the third and fourth conditions in Eq.~(\ref{eq:jz_conditions_3rd_term}) cannot be satisfied. Thus, the term proportional to 
$C^{lpF,K\pi}_{-\alpha\kappa,\lambda-\beta}$ in Eq.~(\ref{eq:cross_term_including_check_C_F}) vanishes for $K\geq 0$ under the imposed convention and, likewise, the term proportional to 
$C^{lpF,K\pi}_{-\beta\lambda,\kappa-\alpha}$ also vanishes for those $K$ values. 
In addition to those two terms, the term proportional to 
$C^{lpF,K\pi}_{\kappa-\alpha,-\beta\lambda}$ and that proportional to 
$C^{lpF,K\pi}_{\lambda-\beta,-\alpha\kappa}$ also vanish for $K=0$. It has been confirmed that those terms are not actually treated in the computation, if they are included in the program. 

\section{\label{app:me_pn_pairing_int} Matrix element of pn pairing interaction}
The matrix element $\bar{V}^\textrm{\scriptsize{pnpair}}_{\mu\nu,\kappa\lambda}$ of the pn pairing interaction in Eq.~(\ref{eq:pn_pairing_interaction}), which is a component of $\bar{V}^\textrm{\scriptsize{pp}}_{K^\prime KL^\prime L}$ in Eq.~(A5) in Appendix A of Ref.~\cite{Ter05}, is given by 
\begin{eqnarray}
\bar{V}^\textrm{\scriptsize{pnpair}}_{\mu\nu,\kappa\lambda} &=& 
\int d^3\bm{r}_1 \int d^3\bm{r}_2 \sum_{\sigma_1\sigma_2}\sum_{\sigma_1^\prime\sigma_2^\prime}
\psi_\mu^\ast(\bm{r}_1\sigma_1 p) \psi_\nu^\ast(\bm{r}_2 \sigma_2 n)
\nonumber \\
&&\times\frac{1}{2}\Big\{ \langle \sigma_1\sigma_2,T=0,T_z=0|g^\textrm{\scriptsize{pnpair}}_{T=0} \delta(\bm{r}_1-\bm{r}_2) P_{S=1} P_{T=0}
\big( |\sigma_1^\prime \sigma_2^\prime \rangle +|\sigma_2^\prime\sigma_1^\prime\rangle \big) 
|T=0,T_z=0\rangle \nonumber \\
&&+\langle \sigma_1\sigma_2,T=1,T_z=0|g^\textrm{\scriptsize{pnpair}}_{T=1} \delta(\bm{r}_1-\bm{r}_2) 
P_{S=0} P_{T=1}
\big( |\sigma_1^\prime \sigma_2^\prime \rangle -|\sigma_2^\prime\sigma_1^\prime\rangle \big) 
|T=1,T_z=0\rangle \Big\} \nonumber \\
&&\times\psi_\kappa(\bm{r}_1\sigma_1^\prime p) \psi_\lambda(\bm{r}_2\sigma_2^\prime n).
\nonumber \\
\label{eq:me_pn_pairing_int_1st_step}
\end{eqnarray}
In this equation, it is assumed that $\mu$ and $\kappa$ are proton, and $\nu$ and $\lambda$ are neutron. The canonical single-particle wave function is expressed as 
\begin{eqnarray}
\psi_\mu(\bm{r}) = \sum_{\sigma\tau}\psi_\mu(\bm{r}\sigma\tau)|\sigma\rangle|\tau\rangle,
\end{eqnarray}
where $\sigma$ and $\tau$ are the $z$-components of the spin and isospin, respectively, and $|\sigma\rangle$ and 
$|\tau\rangle$ are the associated respective eigen wave functions. Using the canonical wave function of the axially symmetric system
\begin{eqnarray}
\psi_\mu(\bm{r}\sigma\tau) =\frac{1}{\sqrt{2\pi}} {\cal F}_\mu(\sigma\tau;z,\rho) e^{i(j^z_\mu-\sigma)\phi}, 
\end{eqnarray} 
with the cylindrical coordinate $(z,\rho,\phi)$, it follows that 
\begin{eqnarray}
\bar{V}^\textrm{\scriptsize{pnpair}}_{\mu\nu,\kappa\lambda} &=&
\delta_{-j^z_\mu-j^z_\nu+j^z_\kappa+j^z_\lambda,0}
\delta_{\pi_\mu \pi_\nu \pi_\kappa \pi_\lambda,1}
\bigg[
g^\textrm{\scriptsize{pnpair}}_{T=0} \frac{1}{2\pi}
\int_0^\infty d\rho\rho \int_0^\infty dz \Big\{
2\sum_\sigma {\cal F}_\mu^\ast(\sigma p;z,\rho) {\cal F}_\nu^\ast(\sigma n;z,\rho) \nonumber\\
&&\times{\cal F}_\kappa(\sigma p;z,\rho) {\cal F}_\lambda(\sigma n;z,\rho)
+{\cal G}^{T=0\ast}_{\mu\nu}(z,\rho) {\cal G}^{T=0}_{\kappa\lambda}(z,\rho) \Big\} \nonumber\\
&&+g^\textrm{\scriptsize{pnpair}}_{T=1} \frac{1}{2\pi} 
\int_0^\infty d\rho \rho \int_0^\infty dz 
{\cal G}^{T=1\ast}_{\mu\nu}(z,\rho) {\cal G}^{T=1}_{\kappa\lambda}(z,\rho) \bigg],
\label{eq:me_pn_pairing_int_2nd_step}
\end{eqnarray}
\begin{eqnarray}
&&{\cal G}^{T=0}_{\mu\nu}(z,\rho) = \sum_\sigma {\cal F}_\mu(\sigma p;z,\rho) {\cal F}_\nu(-\sigma n; z,\rho) , \nonumber\\
&&{\cal G}^{T=1}_{\mu\nu}(z,\rho) = \sum_\sigma {\cal F}_\mu(\sigma p;z,\rho) {\cal F}_\nu(-\sigma n; z,\rho)
(-)^{\frac{1}{2}-\sigma}.
\end{eqnarray}
This equation is used in the numerical calculation, in which ${\cal F}_\mu(\sigma\tau;z,\rho)$ is real.

\begin{acknowledgments}
 This study was supported by the HPCI Strategic Program Field 5, and by a JSPS Grant-in-Aid for Scientific Research (C) under subject number 26400265 and a Grant-in-Aid for Scientific Research on Innovative Areas, Public Research under subject number 15H01029. The following computers were used: the K computer at AICS, RIKEN, through HPCI Strategic Program Field 5 (hp120287); FX100 at ITC, Nagoya University (z48530u); FX10 at ITC, University of Tokyo (f16000); and Coma at CCS, University of Tsukuba, through the Interdisciplinary Computational Science Program in CCS (TKBNDFT).
\end{acknowledgments}
\vfill

\begin{thebibliography}{29}
\expandafter\ifx\csname natexlab\endcsname\relax\def\natexlab#1{#1}\fi
\expandafter\ifx\csname bibnamefont\endcsname\relax
  \def\bibnamefont#1{#1}\fi
\expandafter\ifx\csname bibfnamefont\endcsname\relax
  \def\bibfnamefont#1{#1}\fi
\expandafter\ifx\csname citenamefont\endcsname\relax
  \def\citenamefont#1{#1}\fi
\expandafter\ifx\csname url\endcsname\relax
  \def\url#1{\texttt{#1}}\fi
\expandafter\ifx\csname urlprefix\endcsname\relax\def\urlprefix{URL }\fi
\providecommand{\bibinfo}[2]{#2}
\providecommand{\eprint}[2][]{\url{#2}}

\bibitem[{\citenamefont{Umehara et~al.}(2008)}]{Ume08}
\bibinfo{author}{\bibfnamefont{S.}~\bibnamefont{Umehara}} \bibnamefont{et~al.},
  \bibinfo{journal}{Phys. Rev. C} \textbf{\bibinfo{volume}{78}},
  \bibinfo{pages}{058501} (\bibinfo{year}{2008}).

\bibitem[{\citenamefont{Arnaboldi et~al.}(2008)}]{Arn08}
\bibinfo{author}{\bibfnamefont{C.}~\bibnamefont{Arnaboldi}}
  \bibnamefont{et~al.}, \bibinfo{journal}{Phys. Rev. C}
  \textbf{\bibinfo{volume}{78}}, \bibinfo{pages}{035502}
  (\bibinfo{year}{2008}).

\bibitem[{\citenamefont{\mbox{EXO-200 Collaboration}}(2014)}]{EXO14}
\bibinfo{author}{\bibnamefont{\mbox{EXO-200 Collaboration}}},
  \bibinfo{journal}{Nature} \textbf{\bibinfo{volume}{510}},
  \bibinfo{pages}{229} (\bibinfo{year}{2014}).

\bibitem[{\citenamefont{Ackermann et~al.}(2013)}]{Ack13}
\bibinfo{author}{\bibfnamefont{K.-H.} \bibnamefont{Ackermann}}
  \bibnamefont{et~al.}, \bibinfo{journal}{Eur. Phys. J. C}
  \textbf{\bibinfo{volume}{73}}, \bibinfo{pages}{2330} (\bibinfo{year}{2013}).

\bibitem[{\citenamefont{Gando et~al.}(2013)}]{Gan13}
\bibinfo{author}{\bibfnamefont{A.}~\bibnamefont{Gando}} \bibnamefont{et~al.},
  \bibinfo{journal}{Phys. Rev. Lett.} \textbf{\bibinfo{volume}{110}},
  \bibinfo{pages}{062502} (\bibinfo{year}{2013}).

\bibitem[{\citenamefont{Faessler}(2012)}]{Fae12J}
\bibinfo{author}{\bibfnamefont{A.}~\bibnamefont{Faessler}},
  \bibinfo{journal}{Jour. Phys.: Conf. Ser.} \textbf{\bibinfo{volume}{337}},
  \bibinfo{pages}{012065} (\bibinfo{year}{2012}).

\bibitem[{\citenamefont{Vergados et~al.}(2012)\citenamefont{Vergados, Ejiri,
  and {\v S}imkovic}}]{Ver12}
\bibinfo{author}{\bibfnamefont{J.~D.} \bibnamefont{Vergados}},
  \bibinfo{author}{\bibfnamefont{H.}~\bibnamefont{Ejiri}}, \bibnamefont{and}
  \bibinfo{author}{\bibfnamefont{F.}~\bibnamefont{{\v S}imkovic}},
  \bibinfo{journal}{Rep. Prog. Phys.} \textbf{\bibinfo{volume}{75}},
  \bibinfo{pages}{106301} (\bibinfo{year}{2012}).

\bibitem[{\citenamefont{Raduta et~al.}(1993)\citenamefont{Raduta, Faessler, and
  Delion}}]{Rad93}
\bibinfo{author}{\bibfnamefont{A.~A.} \bibnamefont{Raduta}},
  \bibinfo{author}{\bibfnamefont{A.}~\bibnamefont{Faessler}}, \bibnamefont{and}
  \bibinfo{author}{\bibfnamefont{D.~S.} \bibnamefont{Delion}},
  \bibinfo{journal}{Nucl. Phys. A} \textbf{\bibinfo{volume}{564}},
  \bibinfo{pages}{185} (\bibinfo{year}{1993}).

\bibitem[{\citenamefont{{\v S}imkovic et~al.}(2008)\citenamefont{{\v S}imkovic,
  Faessler, Rodin, Vogel, and Engel}}]{Sim08}
\bibinfo{author}{\bibfnamefont{F.}~\bibnamefont{{\v S}imkovic}},
  \bibinfo{author}{\bibfnamefont{A.}~\bibnamefont{Faessler}},
  \bibinfo{author}{\bibfnamefont{V.~A.} \bibnamefont{Rodin}},
  \bibinfo{author}{\bibfnamefont{P.}~\bibnamefont{Vogel}}, \bibnamefont{and}
  \bibinfo{author}{\bibfnamefont{J.}~\bibnamefont{Engel}},
  \bibinfo{journal}{Phys. Rev. C} \textbf{\bibinfo{volume}{77}},
  \bibinfo{pages}{045503} (\bibinfo{year}{2008}).

\bibitem[{\citenamefont{Rodin and Faessler}(2002)}]{Rod02}
\bibinfo{author}{\bibfnamefont{V.}~\bibnamefont{Rodin}} \bibnamefont{and}
  \bibinfo{author}{\bibfnamefont{A.}~\bibnamefont{Faessler}},
  \bibinfo{journal}{Phys. Rev. C} \textbf{\bibinfo{volume}{66}},
  \bibinfo{pages}{051303(R)} (\bibinfo{year}{2002}).

\bibitem[{\citenamefont{Pacearescu et~al.}(2003)\citenamefont{Pacearescu,
  Rodin, \v{S}imkovic, and Faessler}}]{Pac03}
\bibinfo{author}{\bibfnamefont{L.}~\bibnamefont{Pacearescu}},
  \bibinfo{author}{\bibfnamefont{V.}~\bibnamefont{Rodin}},
  \bibinfo{author}{\bibfnamefont{F.}~\bibnamefont{\v{S}imkovic}},
  \bibnamefont{and} \bibinfo{author}{\bibfnamefont{A.}~\bibnamefont{Faessler}},
  \bibinfo{journal}{Phys. Rev. C} \textbf{\bibinfo{volume}{68}},
  \bibinfo{pages}{064310} (\bibinfo{year}{2003}).

\bibitem[{\citenamefont{Kortelainen and Suhonen}(2007)}]{Kor07}
\bibinfo{author}{\bibfnamefont{M.}~\bibnamefont{Kortelainen}} \bibnamefont{and}
  \bibinfo{author}{\bibfnamefont{J.}~\bibnamefont{Suhonen}},
  \bibinfo{journal}{Phys. Rev. C} \textbf{\bibinfo{volume}{76}},
  \bibinfo{pages}{024315} (\bibinfo{year}{2007}).

\bibitem[{\citenamefont{Fang et~al.}(2015)\citenamefont{Fang, Faessler, and {\v
  S}imkovic}}]{Fan15}
\bibinfo{author}{\bibfnamefont{D.}~\bibnamefont{Fang}},
  \bibinfo{author}{\bibfnamefont{A.}~\bibnamefont{Faessler}}, \bibnamefont{and}
  \bibinfo{author}{\bibfnamefont{F.}~\bibnamefont{{\v S}imkovic}},
  \bibinfo{journal}{Phys. Rev. C} \textbf{\bibinfo{volume}{92}},
  \bibinfo{pages}{044301} (\bibinfo{year}{2015}).

\bibitem[{\citenamefont{Terasaki}(2013)}]{Ter13}
\bibinfo{author}{\bibfnamefont{J.}~\bibnamefont{Terasaki}},
  \bibinfo{journal}{Phys. Rev. C} \textbf{\bibinfo{volume}{87}},
  \bibinfo{pages}{024316} (\bibinfo{year}{2013}).

\bibitem[{\citenamefont{Terasaki}(2015)}]{Ter15}
\bibinfo{author}{\bibfnamefont{J.}~\bibnamefont{Terasaki}},
  \bibinfo{journal}{Phys. Rev. C} \textbf{\bibinfo{volume}{91}},
  \bibinfo{pages}{034318} (\bibinfo{year}{2015}).

\bibitem[{\citenamefont{Fang et~al.}(2011)\citenamefont{Fang, Faessler, Rodin,
  and \v{S}imkovic}}]{Fan11}
\bibinfo{author}{\bibfnamefont{D.-L.} \bibnamefont{Fang}},
  \bibinfo{author}{\bibfnamefont{A.}~\bibnamefont{Faessler}},
  \bibinfo{author}{\bibfnamefont{V.}~\bibnamefont{Rodin}}, \bibnamefont{and}
  \bibinfo{author}{\bibfnamefont{F.}~\bibnamefont{\v{S}imkovic}},
  \bibinfo{journal}{Phys. Rev. C} \textbf{\bibinfo{volume}{83}},
  \bibinfo{pages}{034320} (\bibinfo{year}{2011}).

\bibitem[{\citenamefont{Vogel}(2010)}]{Vog10}
\bibinfo{author}{\bibfnamefont{P.}~\bibnamefont{Vogel}},
  \emph{\bibinfo{title}{Current Aspects of Neutrino Physics, P.O. Caldwell,
  ed.}} (\bibinfo{publisher}{Springer-Verlag}, \bibinfo{address}{New York},
  \bibinfo{year}{2010}), p. \bibinfo{pages}{177}.

\bibitem[{\citenamefont{Brown et~al.}(2014)\citenamefont{Brown, Horoi, and
  Sen'kov}}]{Bro14}
\bibinfo{author}{\bibfnamefont{B.~A.} \bibnamefont{Brown}},
  \bibinfo{author}{\bibfnamefont{M.}~\bibnamefont{Horoi}}, \bibnamefont{and}
  \bibinfo{author}{\bibfnamefont{R.~A.} \bibnamefont{Sen'kov}},
  \bibinfo{journal}{Phys. Rev. Lett.} \textbf{\bibinfo{volume}{113}},
  \bibinfo{pages}{262501} (\bibinfo{year}{2014}).

\bibitem[{\citenamefont{Doi et~al.}(1985)\citenamefont{Doi, Kotani, and
  Takasugi}}]{Doi85}
\bibinfo{author}{\bibfnamefont{M.}~\bibnamefont{Doi}},
  \bibinfo{author}{\bibfnamefont{T.}~\bibnamefont{Kotani}}, \bibnamefont{and}
  \bibinfo{author}{\bibfnamefont{E.}~\bibnamefont{Takasugi}},
  \bibinfo{journal}{Prog. Theor. Phys. Suppl.} \textbf{\bibinfo{volume}{83}},
  \bibinfo{pages}{1} (\bibinfo{year}{1985}).

\bibitem[{\citenamefont{Ullah and Gupta}(1972)}]{Ull72}
\bibinfo{author}{\bibfnamefont{N.}~\bibnamefont{Ullah}} \bibnamefont{and}
  \bibinfo{author}{\bibfnamefont{K.~K.} \bibnamefont{Gupta}},
  \bibinfo{journal}{Nucl. Phys.} \textbf{\bibinfo{volume}{A186}},
  \bibinfo{pages}{331} (\bibinfo{year}{1972}).

\bibitem[{\citenamefont{Terasaki et~al.}(2005)\citenamefont{Terasaki, Engel,
  Bender, Dobaczewski, Nazarewicz, and Stoitsov}}]{Ter05}
\bibinfo{author}{\bibfnamefont{J.}~\bibnamefont{Terasaki}},
  \bibinfo{author}{\bibfnamefont{J.}~\bibnamefont{Engel}},
  \bibinfo{author}{\bibfnamefont{M.}~\bibnamefont{Bender}},
  \bibinfo{author}{\bibfnamefont{J.}~\bibnamefont{Dobaczewski}},
  \bibinfo{author}{\bibfnamefont{W.}~\bibnamefont{Nazarewicz}},
  \bibnamefont{and} \bibinfo{author}{\bibfnamefont{M.}~\bibnamefont{Stoitsov}},
  \bibinfo{journal}{Phys. Rev. C} \textbf{\bibinfo{volume}{71}},
  \bibinfo{pages}{034310} (\bibinfo{year}{2005}).

\bibitem[{\citenamefont{Terasaki and Engel}(2010)}]{Ter10}
\bibinfo{author}{\bibfnamefont{J.}~\bibnamefont{Terasaki}} \bibnamefont{and}
  \bibinfo{author}{\bibfnamefont{J.}~\bibnamefont{Engel}},
  \bibinfo{journal}{Phys. Rev. C} \textbf{\bibinfo{volume}{82}},
  \bibinfo{pages}{034326} (\bibinfo{year}{2010}).

\bibitem[{\citenamefont{Bender et~al.}(2002)\citenamefont{Bender, Dobaczewski,
  Engel, and Nazarewicz}}]{Ben02}
\bibinfo{author}{\bibfnamefont{M.}~\bibnamefont{Bender}},
  \bibinfo{author}{\bibfnamefont{J.}~\bibnamefont{Dobaczewski}},
  \bibinfo{author}{\bibfnamefont{J.}~\bibnamefont{Engel}}, \bibnamefont{and}
  \bibinfo{author}{\bibfnamefont{W.}~\bibnamefont{Nazarewicz}},
  \bibinfo{journal}{Phys. Rev. C} \textbf{\bibinfo{volume}{65}},
  \bibinfo{pages}{054322} (\bibinfo{year}{2002}).

\bibitem[{\citenamefont{Mustonen and Engel}(2013)}]{Mus13}
\bibinfo{author}{\bibfnamefont{M.~T.} \bibnamefont{Mustonen}} \bibnamefont{and}
  \bibinfo{author}{\bibfnamefont{J.}~\bibnamefont{Engel}},
  \bibinfo{journal}{Phys. Rev} \textbf{\bibinfo{volume}{87}},
  \bibinfo{pages}{064302} (\bibinfo{year}{2013}).

\bibitem[{\citenamefont{Barabash}(2013)}]{Bar13}
\bibinfo{author}{\bibfnamefont{A.~S.} \bibnamefont{Barabash}},
  \emph{\bibinfo{title}{Workshop on Calculation of Double-beta-decay Matrix
  Elements (MEDEX'13), O. Civitarese, I. Stekl, and J. Suhonen eds.}}
  (\bibinfo{publisher}{AIP Publishing}, \bibinfo{address}{Melville},
  \bibinfo{year}{2013}), p.~\bibinfo{pages}{11}.

\bibitem[{\citenamefont{Kotila and Iachello}(2012)}]{Kot12}
\bibinfo{author}{\bibfnamefont{J.}~\bibnamefont{Kotila}} \bibnamefont{and}
  \bibinfo{author}{\bibfnamefont{F.}~\bibnamefont{Iachello}},
  \bibinfo{journal}{Phys. Rev. C} \textbf{\bibinfo{volume}{85}},
  \bibinfo{pages}{034316} (\bibinfo{year}{2012}).

\bibitem[{\citenamefont{Bartel et~al.}(1982)\citenamefont{Bartel, Quentin,
  Brack, Guet, and H{\aa}kansson}}]{Bar82}
\bibinfo{author}{\bibfnamefont{J.}~\bibnamefont{Bartel}},
  \bibinfo{author}{\bibfnamefont{P.}~\bibnamefont{Quentin}},
  \bibinfo{author}{\bibfnamefont{M.}~\bibnamefont{Brack}},
  \bibinfo{author}{\bibfnamefont{C.}~\bibnamefont{Guet}}, \bibnamefont{and}
  \bibinfo{author}{\bibfnamefont{H.-B.} \bibnamefont{H{\aa}kansson}},
  \bibinfo{journal}{Nucl. Phys. A} \textbf{\bibinfo{volume}{386}},
  \bibinfo{pages}{79} (\bibinfo{year}{1982}).

\bibitem[{Sca()}]{Sca12}
\bibinfo{howpublished}{[http://www.netlib.org/scalapack]}.

\bibitem[{\citenamefont{\mbox{Message \ \ \ Passing \ \ \ Interface \ \ \
  Forum}}(2015)}]{mpi15}
\bibinfo{author}{\bibnamefont{\mbox{Message \ \ \ Passing \ \ \ Interface \ \ \
  Forum}}}, \emph{\bibinfo{title}{Mpi: A message-passing interface standard}},
  \bibinfo{howpublished}{[http://www.mpi-forum.org]} (\bibinfo{year}{2015}).

\end{thebibliography}

\end{document}